\documentclass{appolb}%
\usepackage{graphicx}
\usepackage{amsmath}
\usepackage{amsfonts}
\usepackage{amssymb}%
\begin{document}

\title{ A covariant nonlocal Lagrangian for the description of the scalar kaonic sector}
\author{Milena So{\l }tysiak$^{1}$, Francesco Giacosa$^{1,2}$
\address{$^{1}$Institute of
Physics, Jan Kochanowski University, PL-25406 Kielce, Poland}
\address{$^{2}$Institute for Theoretical Physics, Goethe
University, D-60438 Frankfurt am Main, Germany} }
\maketitle

\begin{abstract}
Mesons are extended objects, hence their interaction can be described by
utilizing form factors. At the Lagrangian level, one can use nonlocal
interaction terms. Here we describe two possible nonlocal Lagrangians leading
to a 3D form factor: the first one is simple but does not fulfill covariance
(if one insists on a 3D cutoff), the second extension is more involved but
guarantees covariance. Such form factors are useful when calculating mesonic
loops. As an important example, we discuss the scalar kaonic sector,
$I(J^{P})=\frac{1}{2}(0^{+})$. The Lagrangian contains a single scalar kaon
(the well-establish state $K_{0}^{\ast}(1430)$), but through loops
$K_{0}^{\ast}(800)$ emerges as a dynamically generated companion pole (which
disappears in the large-$N_{c}$ limit).

\end{abstract}

\section{Introduction}

Mesonic Lagrangians are often used to describe the decays and the spectral
functions of resonances listed in the PDG \cite{pdg}. Such Lagrangians make
use of some symmetries of the underlying QCD (such as flavor or chiral
symmetry), e.g. chiral perturbation theory \cite{scherer} and (extended) sigma
models \cite{dick}. In particular, in the case of some enigmatic resonances,
such as the light $K_{0}^{\ast}(800)$ meson addressed in this work, the role
of mesonic loops turns out to be extremely important, see e.g. Refs.
\cite{dullemond,oller,black,lupo,our,fariborz} and refs. therein.

Yet, mesons are not elementary particles but are extended objects with a
radius of about $0.5$ fm. Some type of form factor is needed. Already in the
$^{3}P_{0}$ model, e.g. Ref. \cite{amsler}, a form factor reducing the decay
for increasing phase space is present. This form factor is also useful when
calculating quantum fluctuations (loops), since all contributions become
finite (see Sec. 2). It is possible to include a form factor directly in the
Lagrangian by using nonlocal interaction terms, e.g. Refs.
\cite{nonlocal,lyubopion,lyubods0}. In many studies of mesons, a 3D form
factor is employed for simplicity.\ This is usually regarded as a breaking of
covariance. Here we discuss how to introduce nonlocal terms which deliver the
desired 3D cutoff (Sec. 3). Interestingly, an extension which preserves
covariance is possible \cite{scalarfermion}. Implications for the scalar
kaonic sector and conclusions are discussed in Sec. 4.

\section{The `ad hoc' introduction of a form factor}

In general, effective Lagrangians contain both derivative and nonderivative
interaction terms \cite{scherer,dick}. A prototype of such a Lagrangian in
which a (scalar) state $S$ interacts with two (pseudoscalar) particles
$\varphi_{1}$ and $\varphi_{2}$ reads:
\begin{equation}
\mathcal{L}_{S\varphi_{1}\varphi_{2}}=aS\varphi_{1}\varphi_{2}+bS\partial
_{\mu}\varphi_{1}\partial^{\mu}\varphi_{2}\text{ .}\label{laglocal}%
\end{equation}
The scalar kaonic sector is obtained upon identifying $S=K_{0}^{\ast+}$ (in
first approximation corresponding to $K_{0}^{\ast}(1430)$) and $\varphi
_{1}\equiv\pi^{0}$ and $\varphi_{2}\equiv K^{-}$ (other isospin combinations
are here neglected). The tree-level decay width $S\rightarrow\varphi
_{1}\varphi_{2}$ as function of the `running' mass $m$ of $S$ reads:%
\begin{equation}
\Gamma_{S\rightarrow\varphi_{1}\varphi_{2}}(m)=\frac{\left\vert \vec{k}%
_{1}\right\vert }{8\pi m^{2}}\left[  a-b\frac{m^{2}-m_{1}^{2}-m_{2}^{2}}%
{2}\right]  ^{2}\text{ ,}%
\end{equation}
where $m_{1}$ is the mass of $\varphi_{1}$ and $m_{2}$ the mass of
$\varphi_{2}$. The quantity $\left\vert \vec{k}_{1}\right\vert $ is the
modulus of the three-momentum of (one of the) outgoing particle(s):%
\begin{equation}
\left\vert \vec{k}_{1}\right\vert =\frac{\sqrt{m^{4}+(m_{1}^{2}-m_{2}^{2}%
)^{2}-2m^{2}(m_{1}^{2}+m_{2}^{2})}}{2m}\text{ .}%
\end{equation}
The actual value of the tree-level decay width is obtained by setting $m$ to
the tree-level (nominal) mass $m_{S}$ of the field $S$, $\Gamma_{S\rightarrow
\varphi_{1}\varphi_{2}}(m_{S}).$ The Lagrangian (\ref{laglocal}) as it stands
is not suitable for loop calculations. One can regularize the theory by
considering an `ad hoc' modification of both vertices via a form factor:%
\begin{equation}
a\rightarrow af_{\Lambda}(\vec{k}_{1}^{2})\text{ , }b\rightarrow bf_{\Lambda
}(\vec{k}_{1}^{2})\text{ with }f_{\Lambda}(\vec{k}_{1}^{2})=e^{-\vec{k}%
_{1}^{2}/\Lambda^{2}}\label{abmod}%
\end{equation}
The quantity $\Lambda$ is often called `cutoff' (a smooth cutoff here). The
choice of a Gaussian is part of the modelling and is obviously not unique.
Anyway, there is clear physical motivation behind $\Lambda$: it is the energy
scale which takes into account that mesons are extended objects. The dimension
of the system is roughly given by $1/\Lambda.$ Numerically $\Lambda$ lies
between $0.5$ and $1$ GeV \cite{lupo}. The `local' limit is recovered for
$\Lambda\rightarrow\infty.$ As a consequence of Eq. (\ref{abmod}), the decay
width changes as
\begin{equation}
\Gamma_{S\rightarrow\varphi_{1}\varphi_{2}}(m)\rightarrow\Gamma_{S\rightarrow
\varphi_{1}\varphi_{2}}^{\Lambda}(m)=\Gamma_{S\rightarrow\varphi_{1}%
\varphi_{2}}(m)f_{\Lambda}^{2}(\vec{k}_{1}^{2})\text{ ,}\label{gammalambda}%
\end{equation}
and is similar to the form factor implemented in the $^{3}P_{0}$ model (see
e.g. \cite{amsler} and the `quark model' review in\ Ref. \cite{pdg}).\ 

The contribution of the loop $\Pi(m^{2})$ in which the particles $\varphi_{1}$
and $\varphi_{2}$ circulate as calculated from the original local Lagrangian
(\ref{laglocal}) reads%
\begin{equation}
\Pi(m^{2})=\int\frac{d^{4}k_{1}}{(2\pi)^{4}}\frac{\left[  a-b\left(
k_{1}\cdot k_{2}\right)  \right]  ^{2}}{\left[  k_{1}^{2}-m_{1}^{2}%
+i\varepsilon\right]  \left[  k_{2}^{2}-m_{2}^{2}+i\varepsilon\right]  }\text{
,}%
\end{equation}
where the constraint $k_{2}=p-k_{1}$ is understood and $p$ is the momentum of
the unstable particle $S.$ In its reference frame $p=(m,\vec{0}).$ As
mentioned above, this loop contribution is divergent (with $\Lambda^{4}$). The
substitution (\ref{abmod}) makes it convergent thanks to the form-factor:%
\begin{equation}
\Pi(m^{2})\rightarrow\Pi_{\Lambda}(m^{2})=\int\frac{d^{4}k_{1}}{(2\pi)^{4}%
}\frac{\left[  a-b\left(  k_{1}\cdot k_{2}\right)  \right]  ^{2}f_{\Lambda
}^{2}(\vec{k}_{1}^{2})}{\left[  k_{1}^{2}-m_{1}^{2}+i\varepsilon\right]
\left[  k_{2}^{2}-m_{2}^{2}+i\varepsilon\right]  }\text{ .} \label{looplambda}%
\end{equation}
At this point, one may object that the form factor breaks covariance, since it
depends on the three-momentum only. We will show in the next section that this
is not necessarily the case. Once the form factor is introduced, the full
propagator of the particle $S$ is calculated as $\Delta_{{S}}(p^{2}%
=m^{2})=\left[  m^{2}-M_{0}^{2}+\Pi_{\Lambda}(m^{2})+i\varepsilon\right]
^{-1}$ where $M_{0}$ is the bare mass of $S$ and $\operatorname{Im}%
\Pi_{\Lambda}(m^{2})=m\Gamma_{S\rightarrow\varphi_{1}\varphi_{2}}^{\Lambda
}(m).$ At the one-loop level, the Breit-Wigner mass of $S$ is defined as
$m_{S}^{2}-M_{0}^{2}+\operatorname{Re}\Pi(m_{S}^{2})=0,$ while the pole mass
is obtained by solving the equation $s-M_{0}^{2}+\Pi(s)=0$ in the complex
$s$-plane, see e.g. Ref. \cite{our,a0rev} for the scalar $I=1/2$ and $I=1$ sectors.

The real part of the loop can be also obtained as $\operatorname{Re}%
\Pi_{\Lambda}(s=m^{2})=c+\frac{1}{\pi}PP\int_{m_{1}+m_{2}}^{\infty}%
\frac{\operatorname{Im}\Pi_{\Lambda}(s^{\prime})}{s^{\prime}-s}ds^{\prime}$
where $c$ is an additional constant due to a subtlety of QFT containing
derivatives, see Ref. \cite{a0rev} for details. For $\Lambda\rightarrow\infty
$, $\operatorname{Im}\Pi_{\Lambda}(s^{\prime})$ scale as $s^{\prime2},$ then a
three-time subtracted dispersion relation would be needed. This is quite
cumbersome and reinforces the viewpoint that -for the particular case of
hadronic physics- a physical cutoff is a meaningful procedure.

\section{Nonlocal Lagrangians}

The modification of Eq. (\ref{abmod}) and, consequently, the form factor in
Eqs. (\ref{gammalambda}) and (\ref{looplambda}) have been introduced as an `ad
hoc' change of the Feynman rules.\ Yet, it is possible to modify the
Lagrangian in order that the modified equations automatically follows from it.

\textbf{Nonlocal extension 1}: The easiest way (see Refs.
\cite{nonlocal,lyubopion}) is to consider the following nonlocal Lagrangian
(for simplicity we discuss here only the $a$-term in Eq. (\ref{laglocal}), but
the $b$-term is very similar):%
\begin{equation}
aS\varphi_{1}\varphi_{2}\rightarrow aS\int d^{4}y\varphi_{1}(x+y/2)\varphi
_{2}(x-y/2)\Phi(y)\text{ .}%
\end{equation}
The Feynman rule at the $a$-vertex is modified (upon defining $k_{1}=p/2+q,$
$k_{2}=p/2-q$, hence: $q=(k_{1}-k_{2})/2$):
\begin{equation}
a\rightarrow a\int d^{4}ye^{-iqy}\Phi(y)=a\varphi\left(  q\right)  \text{ .}%
\end{equation}
For $\Phi(y)=\delta^{(4)}(y)$ one reobtains the local limit. A covariant form
factor requires a dependence $\Phi(y^{2})$, hence $\varphi(q^{2})$ follows.
Quite interestingly, if the masses of $\varphi_{1}$ and $\varphi_{2}$ are
equal and on-shell, $q^{2}=-\vec{k}_{1}^{2}$, in agreement with Eq.
(\ref{gammalambda}) upon setting $\varphi(q^{2})=e^{q^{2}/\Lambda^{2}}$.
However, the loop is different from\ Eq. (\ref{looplambda}), since it involve
$\varphi^{2}(q^{2})=\varphi^{2}(4(k_{1}-p)^{2})$ into the integral [and not
simply $f_{\Lambda}^{2}(\vec{k}_{1}^{2})$]. A generalization in case of
unequal masses is discussed in\ Refs. \cite{lyubods0}.

The choice $\Phi(y)=\delta(y^{0})\phi(\vec{y})$ leads to the desired result
$\varphi\left(  q\right)  =f_{\Lambda}(\vec{k}_{1}^{2})$ of Eq. (\ref{abmod})
and also of Eqs. (\ref{gammalambda}) and (\ref{looplambda}), but it explicitly
breaks covariance. This result is anyway valuable since it shows that it is
possible to get the desired 3D cutoff form , but should be used only in the
reference frame of the decaying particle.

\textbf{Nonlocal extension 2}: We aim to determine the Lagrangian which
generates the vertex function (\ref{abmod}) in a covariant manner
\cite{scalarfermion}.\ We start from the general nonlocal expression:%
\begin{equation}
aS\varphi_{1}\varphi_{2}\rightarrow g\int d^{4}zd^{4}y_{1}d^{4}y_{1}%
S(x+z)\varphi_{1}(x+y_{1})\varphi_{2}(x+y_{2})\Phi(z,y_{1},y_{2})\text{ ,}%
\end{equation}
where the vertex-function $\Phi(z,y_{1},y_{2})$ in position space has been
introduced. The case $\Phi(z,y_{1},y_{2})$ $=\delta(z)\delta(y_{1}%
)\delta(y_{2})$ delivers the local limit (\ref{laglocal}). The vertex function
in momentum space is given by:%
\[
\varphi(p,k_{1},k_{2})=\int d^{4}zd^{4}y_{1}d^{4}y_{2}e^{ipz}e^{-ik_{1}y_{1}%
}e^{-ik_{2}y_{2}}\Phi(z,y_{1},y_{2})\text{ .}%
\]
Here we \textit{assume} that $\Phi(z,y_{1},y_{2})$ is such that
\begin{equation}
\varphi(p,k_{1},k_{2})=\varphi\left(  p,q=\frac{k_{1}-k_{2}}{2}\right)
=f_{\Lambda}\left(  \frac{q^{2}p^{2}-(q\cdot p)^{2}}{p^{2}}\right)  \text{ .}%
\end{equation}
It respects covariance because the final form factor is a function of
Lorentz-products. Nevertheless, in the rest frame of $S$ one recovers the
desired dependence:
\[
\frac{q^{2}p^{2}-(q\cdot p)^{2}}{p^{2}}=\vec{k}_{1}^{2}\text{ (for
}p=(m,0)\text{).}%
\]
Note, in order to get the desired expression, $\Phi(z,y_{1},y_{2})=\xi
(z,y_{1}-y_{2})\delta(y_{1}+y_{2}),$ out of which $\varphi(p,k_{1},k_{2})=\int
d^{4}zd^{4}ye^{ipz}e^{-i2qy}\xi(z,y)$ with $y=y_{1}-y_{2}$ and $Y=y_{1}%
+y_{2}.$ This line of reasoning shows that it is -at least in principle-
possible to reconcile covariance with a 3D form factor in the rest frame of
the decaying particle.

\section{Discussions and conclusions}

In this work we have discussed form factors entering in mesonic interaction
terms. We have started from a local Lagrangian and we have modified the
interaction vertex `ad hoc' by introducing a function of the three-momentum of
one outgoing particle. Then, we have presented two nonlocal extensions of the
Lagrangian that deliver the desired expressions without the need of modifying
by hand the Feynman rules. The first nonlocal extension is relatively simple
but delivers the desired form factor at the price of breaking covariance. The
second extension is more involved but delivers a 3D cutoff in the rest frame
of the decaying particle by respecting covariance.

Recently, the expressions of the propagator described in Sec. 2 (which are
then justified by our study of\ Sec. 3) were used to study the positions of
the poles in the complex plane in both the isodoublet- and isovector-scalar
sectors. In both cases, the Lagrangians contains only one seed state,
corresponding to the quark-antiquark resonances $a_{0}(1450)$ and $K_{0}%
^{\ast}(1430),$ respectively. Yet, the loops (with the necessary presence of a
form factor) are strong enough to generate two additional resonances:
$a_{0}(980)$ and $K_{0}^{\ast}(800).$ These are companion poles, and hence
dynamically generated states. Moreover, these states disappear in the
large-$N_{c}$ limit, confirming their non-conventional nature. In particular,
$K_{0}^{\ast}(800)$ is not yet confirmed in the PDG \cite{pdg}. The study
based on the formalism described in these proceedings unequivocally finds a
pole: $(0.745\pm0.029)-i(0.263\pm0.027)$ GeV \cite{our}. This is in agreement
with other works, e.g. Ref. \cite{fariborz} and refs. therein, and reinforces
the need of accepting this state in the PDG.

In both Refs. \cite{our,a0rev} one obtains a similar values of the cutoff:
$\Lambda\simeq0.6$ GeV. Moreover, in\ Ref. \cite{our} different form factors
have been tested. It is found that they cannot fit the pion-kaon scattering
data so well as the Gaussian form factor. Thus, even if there is in principle
no fundamental reason behind a Gaussian function, it nevertheless seems to be
the best choice in order to describe hadronic phenomenology.

\textbf{Acknowledgments:} we thank T.\ Wolkanowski and G. Pagliara for
valuable discussions. F.G. thanks also past conversations on the topic with
Th.\ Gutsche and V. Lyubovitskij.


\begin{thebibliography}{99}                                                                                               %


\bibitem {pdg}K. A. Olive et al. (Particle Data Group), Chin. Phys.
\textbf{C38}, 090001 (2014).

\bibitem {scherer}S.~Scherer,
Adv.\ Nucl.\ Phys.\ \textbf{27} (2003) 277 [arXiv:hep-ph/0210398].


\bibitem {dick}D. Parganlija, P. Kovacs, G. Wolf, F. Giacosa and D. H.
Rischke, Phys. Rev. \textbf{D87}, 014011 (2012) arXiv:1208.0585 [hep-ph];
S.~Janowski, F.~Giacosa and D.~H.~Rischke,
Phys.\ Rev.\ D \textbf{90} (2014) 11, 114005 [arXiv:1408.4921 [hep-ph]].


\bibitem {dullemond}E. van Beveren, T. A. Rijken, K. Metzger, C. Dullemond, G.
Rupp and J. E. Ribeiro, \emph{Z. Phys.} \textbf{C30}, 615-620 (1986)
arXiv:0710.4067 [hep-ph].

\bibitem {oller}
J.~A.~Oller, E.~Oset and J.~R.~Pelaez,
Phys.\ Rev.\ D \textbf{59} (1999) 074001
[hep-ph/9804209].
J.~R.~Pelaez,
Phys.\ Rev.\ Lett.\ \textbf{92} (2004) 102001
[hep-ph/0309292].


\bibitem {black}
D.~Black, A.~H.~Fariborz, F.~Sannino and J.~Schechter,
Phys.\ Rev.\ D \textbf{58} (1998) 054012
[hep-ph/9804273].


\bibitem {lupo}
F.~Giacosa and G.~Pagliara,
Phys.\ Rev.\ C \textbf{76} (2007) 065204
[arXiv:0707.3594 [hep-ph]];
F.~Giacosa and T.~Wolkanowski,
Mod.\ Phys.\ Lett.\ A \textbf{27} (2012) 1250229
[arXiv:1209.2332 [hep-ph]].


\bibitem {our}
T.~Wolkanowski, M.~Soltysiak and F.~Giacosa,
Nucl.\ Phys.\ B \textbf{909} (2016) 418
[arXiv:1512.01071 [hep-ph]].


\bibitem {fariborz}
A.~H.~Fariborz, E.~Pourjafarabadi, S.~Zarepour and S.~M.~Zebarjad,
Phys.\ Rev.\ D \textbf{92} (2015) 113002
[arXiv:1511.01623 [hep-ph]].


\bibitem {amsler}
C.~Amsler and F.~E.~Close,
Phys.\ Rev.\ D \textbf{53} (1996) 295
[hep-ph/9507326].


\bibitem {nonlocal}
Y.~V.~Burdanov, G.~V.~Efimov, S.~N.~Nedelko and S.~A.~Solunin,
Phys.\ Rev.\ D \textbf{54} (1996) 4483
[hep-ph/9601344];
F.~Giacosa, T.~Gutsche and A.~Faessler,
Phys.\ Rev.\ C \textbf{71} (2005) 025202
[hep-ph/0408085];
R.~D.~Bowler and M.~C.~Birse,
Nucl.\ Phys.\ A \textbf{582} (1995) 655
[hep-ph/9407336];
F.~Giacosa, Glueball phenomenology within a nonlocal approach, PhD thesis,
Tuebingen (2105).


\bibitem {lyubopion}
A.~Faessler, T.~Gutsche, M.~A.~Ivanov, V.~E.~Lyubovitskij and P.~Wang,
Phys.\ Rev.\ D \textbf{68} (2003) 014011
[hep-ph/0304031].


\bibitem {lyubods0}
A.~Faessler, T.~Gutsche, V.~E.~Lyubovitskij and Y.~L.~Ma,
Phys.\ Rev.\ D \textbf{76} (2007) 014005
[arXiv:0705.0254 [hep-ph]].
A.~Faessler, T.~Gutsche, V.~E.~Lyubovitskij and Y.~L.~Ma,
Phys.\ Rev.\ D \textbf{76} (2007) 114008
[arXiv:0709.3946 [hep-ph]].


\bibitem {scalarfermion}
F.~Giacosa and G.~Pagliara,
Phys.\ Rev.\ D \textbf{88} (2013) no.2, 025010
[arXiv:1210.4192 [hep-ph]].


\bibitem {a0rev}
T.~Wolkanowski, F.~Giacosa and D.~H.~Rischke,
Phys.\ Rev.\ D \textbf{93} (2016) no.1, 014002
[arXiv:1508.00372 [hep-ph]].

\end{thebibliography}
\end{document}